\documentclass[10pt]{article}

\usepackage{amsmath,amsfonts,amssymb}
\usepackage{latexsym}
\usepackage{float}
\usepackage{xspace}
\usepackage{theorem,times}

\newcommand{\ALCQIR}{\mathcal{ALCQIR}}

\newcommand{\N}{\mathbb{N}}
\newcommand{\CS}{c.s.\ }

\newcommand{\qed}{{\unskip\nobreak\hfil\penalty50
   \hskip2em\hbox{}\nobreak\hfil
   \qedd
   \parfillskip=0pt \finalhyphendemerits=0
    \medskip\goodbreak\noindent}}

\newcommand{\qedd}{\vrule height4pt width 4pt depth0pt}

\newcommand{\modex}[2]{\ensuremath{\langle #1 \rangle_{#2}}}
\newcommand{\modall}[2]{[#1]_{#2}}
\newcommand{\modleq}[2]{\ensuremath{\langle #1 \rangle_{\leq #2}}}
\newcommand{\modgeq}[2]{\ensuremath{\langle #1 \rangle_{\geq #2}}}

\newcommand{\alcq}{\ensuremath{\mathcal{ALCQ}}\xspace}
\newcommand{\sub}{\ensuremath{\textit{clos}}}

\newcommand{\nneg}{\ensuremath{\mathord{\sim}}}

\newcommand{\pspace}{\textsc{PSpace}\xspace}

\newcommand{\exptime}{\textsc{ExpTime}\xspace}

\newcommand{\satgrkr}{\ensuremath{\textsc{SAT}(\grkr)\xspace}}
\newcommand{\satgrkrri}{\ensuremath{\textsc{SAT}(\grkrri)\xspace}}
\newcommand{\satkr}{\ensuremath{\textsc{SAT}(\kr)\xspace}}

\newcommand{\grkr}{\ensuremath{\mathbf{Gr}(\mathbf{K}_\mathcal{R})}\xspace}
\newcommand{\grkrri}{\ensuremath{\mathbf{Gr}(\mathbf{K}_{\mathcal{R}^{-1}_\cap})}\xspace}
\newcommand{\kr}{\ensuremath{\mathbf{K}_\mathcal{R}}\xspace}

\newcommand{\M}{\ensuremath{\mathfrak{M}}}

\newcommand{\ruleand}{\ensuremath{\rightarrow_\wedge}}
\newcommand{\ruleor}{\ensuremath{\rightarrow_\vee}}
\newcommand{\rulegt}{\ensuremath{\rightarrow_>}}
\newcommand{\ruleleq}{\ensuremath{\rightarrow_\leq}}
\newcommand{\rulegeq}{\ensuremath{\rightarrow_\geq}}

\newcommand{\ruleleqz}{\ensuremath{\rightarrow_{\leq 0}}}
\newcommand{\rulechoose}{\ensuremath{\rightarrow_{\text{choose}}}}

\newcommand{\alc}{\ensuremath{\mathcal{ALC}}\xspace}
\newcommand{\Lab}{\ensuremath{\mathcal{L}}\xspace}

\title{\pspace Reasoning for Graded Modal Logics\thanks{This papers appeared in the
    Journal of Logic and Computation, Vol. 10 No. 99-47, pp. 1--22 2000. }}

\author{Stephan Tobies\\
  LuFg Theoretical Computer Science,  RWTH Aachen, \\
  Theoretische Informatik, Ahornstr. 55, D-52074 Aachen, Germany\\
  \texttt{tobies@informatik.rwth-aachen.de}}

\date{}

\renewenvironment{abstract}{\section*{Abstract}\small}{}
\theoremstyle{break}
\theoremheaderfont{\scshape}
\theorembodyfont{\rmfamily}
\newtheorem{definition}{Definition}[section]

\newtheorem{theorem}[definition]{Theorem}

\newtheorem{lemma}[definition]{Lemma}

\newtheorem{corollary}[definition]{Corollary}
\makeatletter

\theoremstyle{plain}
\newtheorem{proof}{Proof.}

\begin{document}

\maketitle

\begin{abstract}
  We present a \pspace algorithm that decides satisfiability of the graded
  modal logic \grkr---a natural extension of propositional modal logic \kr by
  counting expressions---which plays an important role in the area of
  knowledge representation. The algorithm employs a tableaux approach and is
  the first known algorithm which meets the lower bound for the complexity of
  the problem.  Thus, we exactly fix the complexity of the problem and refute
  a \exptime-hardness conjecture. We extend the results to the logic \grkrri,
  which augments \grkr with inverse relations and intersection of
  accessibility relations.  This establishes a kind of ``theoretical
  benchmark'' that all algorithmic approaches can be measured against.

  \vspace{2ex}

  
  \setlength{\unitlength}{1cm}
  \begin{picture}(0,0)
    \put(-2,11){\small \copyright{} Oxford University Press}
    \put(-2,10.6){ \texttt{\small http://www.oup.co.uk/logcom/}}
  \end{picture}
\end{abstract}

\section{Introduction}

Propositional modal logics have found applications in many areas of computer
science.  Especially in the area of knowledge representation, the description
logic (DL) \alc, which is a syntactical variant of the propositional
(multi-)modal logic \kr \cite{Schild91a}, forms the basis of a large number of
formalisms used to represent and reason about conceptual and taxonomical
knowledge of the application domain.  The graded modal logic \grkr extends \kr
by \emph{graded modalities} \cite{Fine72}, i.e., counting expressions which
allow one to express statements of the form ``there are at least (at most) $n$
accessible worlds that satisfy\dots''.  This is especially useful in knowledge
representation because (a) humans tend to describe objects by the number of
other objects they are related to (a stressed person is a person given at least
three assignments that are urgent), and (b) qualifying number restrictions (the
DL's analogue for graded modalities \cite{HollunderBaader-KR-91}) are necessary
for modeling semantic data models~\cite{CLN94}.

\kr is decidable in \pspace and can be embedded into a decidable fragment of
predicate logic \cite{AndrekaVanBenthemNemeti:JPL:98}. Hence, there are two
general approaches for reasoning with \kr: dedicated decision procedures
\cite{SICOMP::Ladner1977,Schmidt-Schauss91,Giunchiglia96a}, and the translation
into first order logic followed by the application of an existing first order
theorem prover \cite{JLOGC::OhlbachS1997,Schmidt97a}.  To compete with the
dedicated algorithms, the second approach has to yield a decision procedure and
it has to be efficient, because the dedicated algorithms usually have optimal
worst-case complexity. For \kr, the first issue is solved and, regarding the
complexity, experimental results show that the algorithm competes well with
dedicated algorithms~\cite{Hustadt97b}. Since experimental result can only be
partially satisfactory, a theoretical complexity result would be desirable, but
there are no exact results on the complexity of the theorem prover approach.

The situation for \grkr is more complicated: \grkr is known to be decidable, but
this result is rather recent~\cite{HollunderBaader-KR-91}, and the known \pspace
upper complexity bound for \grkr is only valid if we assume unary coding of
numbers in the input, which is an unnatural restriction.  For binary coding no
upper bound is known and the problem has been conjectured to be
\exptime-hard~\cite{JLOGC::HoekR1995}. This coincides with the observation that
a straightforward adaptation of the translation technique leads to an exponential
blow-up in the size of the first order formula. This is because it is possible
to store the number $n$ in $\log_k n$-bits if numbers are represented in $k$-ary
coding.  In \cite{OhlbachSchmidtHustadt96} a translation technique that
overcomes this problem is proposed, but a decision procedure for the target
fragment of first order logic yet has to be developed.

In this work we show that reasoning for \grkr is not harder than reasoning for
\kr by presenting an algorithm that decides satisfiability in \pspace, even if
the numbers in the input are binary coded. It is based on the tableaux
algorithms for \kr and tries to prove the satisfiability of a given formula by
explicitly constructing a model for it.  When trying to
generalise the tableaux algorithms for \kr to deal with \grkr, there are some
difficulties: (1) the straightforward approach leads to an incorrect algorithm;
(2) even if this pitfall is avoided, special care has to be taken in order to
obtain a space-efficient solution. As an example for (1), we will show that the
algorithm presented in \cite{JLOGC::HoekR1995} to decide satisfiability of \grkr
is incorrect.  Nevertheless, this algorithm will be the basis of our further
considerations.  Problem (2) is due to the fact that tableaux algorithms try to
prove the satisfiability of a formula by explicitly building a model for it. If
the tested formula requires the existence of $n$ accessible worlds, a tableaux
algorithm will include them in the model it constructs, which leads to
exponential space consumption, at least if the numbers in the input are not
unarily coded or memory is not re-used.  An example for a correct algorithm which
suffers from this problem can be found in~\cite{HollunderBaader-KR-91} and is
briefly presented in this paper. Our algorithm overcomes this problem by
organising the search for a model in a way that allows for the re-use of space
\emph{for each successor}, thus being capable of deciding satisfiability of
\grkr in \pspace.

Using an extension of these techniques we obtain a \pspace algorithm for the logic \grkrri,
which extends \grkr by inverse relations and intersection of relations. This
solves an open problem from~\cite{IC::DoniniLNN1997}.

This paper is an significantly extended and improved version
of~\cite{Tobies-CADE-99}.



\section{Preliminaries}
 
In this section we introduce the graded modal logic \grkr, the extension of the
multi-modal logic \kr with graded modalities,  first introduced in
\cite{Fine72}.

\begin{definition}[Syntax and Semantics of \grkr]
  \label{def:grkn-syntax+semantics}
  Let $\mathcal{P} = \{p_0, p_1, \dots \}$ be a set of propositional atoms and
  $\mathcal{R}$ a set of \emph{relation names}.  The set of
  \grkr-\emph{formulae} is built according to the following rules:
  \begin{enumerate}
  \item every propositional atom is a \grkr-formula, and
  \item if $\phi,\psi_1,\psi_2$ are \grkr-formulae, $n \in \N$, and $R$ is a
    relation name, then $\neg \phi$, $\psi_1 \wedge \psi_2$, $\psi_1 \vee
    \psi_2$, $\modex R n \phi$, and $\modall R n \phi$ are formulae.
  \end{enumerate}
                        
  \noindent The semantics of \grkr-formulae is based on \emph{Kripke structures}
  \[
  \M = (W^\M, \{ R^\M \mid R \in \mathcal{R} \}, V^\M),
  \]
  where $W^\M$ is a non-empty set of worlds, each $R^\M \subseteq W^\M \times W^\M$
  is an \emph{accessibility relation} on worlds (for $R \in \mathcal{R}$), and
  $V^\M$ is a \emph{valuation} assigning subsets of $W^\M$ to the propositional
  atoms in $\mathcal{P}$. For a Kripke structure $\M$, an element $x \in W^\M$,
  and a \grkr-formula, the model relation $\models$ is defined inductively on
  the structure of formulae:
  \begin{align*}
    \M,x & \models p \ \text{iff} \ x \in V^\M(p) \ \text{for} \ p \in \mathcal{P} \\
    \M,x & \models \neg \phi \ \text{iff} \ \M,x \not\models \phi\\
    \M,x & \models \psi_1 \wedge \psi_2 \ \text{iff} \ \M,x \models \psi_1 \
    \text{and} \ \ \M,x \models \psi_2\\
    \M,x & \models \psi_1 \vee \psi_2 \ \text{iff} \ \M,x \models \psi_1 \ 
    \text{or} \ \
    \M,x \models \psi_2 \\
    \M,x & \models \modex R n \phi \ \text{iff} \ \sharp
    R^\M(x,\phi) > n\\
    \M,x & \models \modall R n \phi \ \text{iff} \ \sharp R^\M(x,\neg\phi) \leq
    n
  \end{align*}
  where $\sharp R^\M(x,\phi) := | \{ y \in W^\M \mid (x,y) \in R^\M \ \text{and}
  \ \M,y \models \phi \}|$
  
  The propositional modal logic \kr is defined as the fragment of \grkr in which
  for all modal operators $n = 0$ holds.

  A formula is called \emph{satisfiable} iff there exists a structure
  $\M$ and a world $x \in W^\M$ such that $\M,x \models \phi$.

  By \satkr and \satgrkr we denote the sets of satisfiable formulae of \kr
  and \grkr, respectively.
\end{definition}

As usual, the modal operators $\modex R n$ and $\modall R n$ are dual:
$\sharp R^\M(x,\phi) > n$ means that in $\M$ more than $n$ $R$-successors of
$x$ satisfy $\phi$; $\sharp R^\M(x,\neg \phi) \leq n$ means that in $\M$ all
but at most $n$ $R$-successors satisfy $\phi$.

In the following we will only consider formulae in \emph{negation normal form}
(NNF), a form in which negations have been pushed inwards and  occur in
front of propositional atoms only. We will denote the NNF of $\neg \phi$ by $\nneg
\phi$. The NNF can always be generated in linear time and space by successively
applying the following equivalences from left to right:
\begin{alignat*}{2}
  \neg(\psi_1 \wedge \psi_2) & \equiv \neg \psi_1 \vee \neg \psi_2 & \qquad \neg
  \modex R n \psi & \equiv \modall R n \neg\psi \\
  \neg(\psi_1 \vee
  \psi_2) & \equiv \neg \psi_1 \wedge \neg \psi_2 & \qquad \neg\modall R n
  \psi & \equiv \modex R n \neg\psi 
\end{alignat*}



\section{Reasoning for \grkr}
\label{sec:reasoning}

Before we present our algorithm for deciding satisfiability of \grkr, for
historic and didactic reasons, we present two other solutions: an incorrect one
\cite{JLOGC::HoekR1995}, and a solution that is less
efficient~\cite{HollunderBaader-KR-91}.

From the fact that \satkr{} is \pspace-complete
\cite{SICOMP::Ladner1977,HalpernMoses92}, it immediately follows, that \satgrkr{}
is \pspace-hard.  The algorithms we will consider decide the satisfiability of a
given formula $\phi$ by trying to construct a model for $\phi$.

\subsection{An incorrect algorithm}
\label{sec:wrong-solution}

In~\cite{JLOGC::HoekR1995}, an algorithm for deciding \satgrkr{} is given, which,
unfortunately, is incorrect.
Nevertheless, it will be the basis for our further considerations and thus it is
presented here. It will be referred to as the \emph{incorrect} algorithm. It is
based on an algorithm given in~\cite{IC::DoniniLNN1997} to decide the
satisfiability of the DL $\mathcal{ALCNR}$, which basically is
the restriction of \grkr, where, in formulae of the form $\modex R n \phi$ or
$\modall R n \phi$ with $n>0$, necessarily $\phi = p \vee \neg p$ holds.

The algorithm for \grkr tries to build a model for a formula $\phi$ by
manipulating sets of constraints with the help of so-called \emph{completion
  rules}.  This is a well-known technique to check the satisfiability of modal
formulae, which has already been used to prove decidability and complexity
results for other DLs (e. g.,
\cite{Schmidt-Schauss91,HollunderBaader-KR-91,Baader96a}). These algorithms can
be understood as variants of tableaux algorithms which are used, for example, to
decide satisfiability of the modal logics \kr, $\mathbf{T}_\mathcal{R}$, or
$\mathbf{S4}_\mathcal{R}$ in~\cite{HalpernMoses92}.

\begin{definition}\label{def:constrain-system}
  Let $\mathcal{V}$ be a set of variables. A \emph{constraint system} (c.s.) $S$ is a
  finite set of expressions of the form `$x \models \phi$' and `$Rxy$', where
  $\phi$ is a formula, $R \in \mathcal{R}$, and $x,y \in \mathcal{V}$.
  
  For a \CS $S$, let $\sharp R^S(x,\phi)$ be the number of variables $y$ for
  which $\{R x y, y \models \phi\} \subseteq S$.  The \CS $[z/y]S$ is obtained
  from $S$ by replacing every occurrence of $y$ by $z$; this replacement is said
  to be \emph{safe} iff, for every variable $x$, formula $\phi$, and relation
  symbol $R$ with $\{x \models \modex R n \phi, Rxy, Rxz\} \subseteq S$ we have
  $ \sharp R^{[z/y]S}(x,\phi) > n$.
  
  A \CS $S$ is said to contain a \emph{clash}, iff for a propositional atom $p$,
  a formula $\phi$, and $m \leq n$:
  \[
  \{ x \models p, x \models \neg p \} \subseteq S \ \text{or} \ \{ x \models
  \modex R m \phi, x \models \modall R n \nneg \phi \} \subseteq S.
  \]
  Otherwise it is called \emph{clash-free}.  A \CS $S$ is called \emph{complete}
  iff none of the rules given in Fig.~\ref{fig:expansion-rules-derijke} is
  applicable to $S$.
\end{definition}

To test the satisfiability of a formula $\phi$, the incorrect algorithm works
as follows: it starts with the \CS $ \{ x \models \phi \}$ and successively
applies the rules given in Fig.~\ref{fig:expansion-rules-derijke}, stopping if
a clash is occurs. Both the rule to apply and the formula to add (in the
\ruleor-rule) or the variables to identify (in the \ruleleq-rule) are selected
non-deterministically. The algorithm answers ``$\phi$ is satisfiable'' iff the
rules can be applied in a way that yields a complete and clash-free c.s. The
notion of \emph{safe} replacement of variables is needed to ensure the
termination of the rule application \cite{HollunderBaader-KR-91}.

Since we are interested in \pspace algorithms, non-determinism imposes no
problem due to Savitch's Theorem, which states that deterministic and
non-deterministic polynomial space coincide \cite{JCSS::Savitch1970}.

\begin{figure}
  \begin{center}
    \begin{tabular}{@{ }l@{ }l@{ }l@{}}
      $\ruleand$-rule: & if \hfill 1. & $x \models \psi_1 \wedge \psi_2 \in S$ and \\
      & \hfill 2. & $\{x \models \psi_1, x\models \psi_2\}  \not\subseteq S$ \\ 
      & then & $S \ruleand S \cup \{x \models \psi_1, x \models \psi_2 \}$\\[1ex]
      $\ruleor$-rule: & if \hfill 1. & $(x \models \psi_1 \vee \psi_2) \in S$
      and \\
      & \hfill 2. & $\{x \models \psi_1, x \models \psi_2\} \cap S =
      \emptyset$\\
      & then &  $S \ruleor S \cup \{ x \models \chi \}$ where $\chi \in \{
      \psi_1,\psi_2\}$\\[1ex]
      $\rulegt$-rule: & if \hfill 1. & $x \models \modex R n \phi \in S$ and\\
      & \hfill 2. & $\sharp R^S(x,\phi) \leq n$ \\
      & then & $S \rulegt S \cup \{ Rxy, y \models \phi \}$ where $y$ is a fresh
      variable.\\[1ex]
      $\ruleleqz$-rule: & if \hfill 1. & $x \models \modall R 0 \phi, Rxy \in S$
      and\\
      & \hfill 2. & $y \models \phi \not\in S$\\
      & then & $S \ruleleqz S \cup \{ y \models \phi \}$\\[1ex]
      $\ruleleq$-rule: & if \hfill 1. & $x \models \modall R n \phi, \sharp
      R^S(x,\phi) > n > 0$ and \\
      & \hfill 2. & $Rxy, Rxz  \in S$ and\\
      & \hfill 3.& replacing $y$ by $z$ is safe in $S$\\
      & then & $S \ruleleq [z/y]S$
    \end{tabular}
    \caption{The incorrect completion rules for \grkr.}
    \label{fig:expansion-rules-derijke}
  \end{center}
\end{figure}

To prove the correctness of a non-deterministic completion algorithm, it is
sufficient to prove three properties of the model generation process:
\begin{enumerate}
\item Termination: Any sequence of rule applications is finite.
\item Soundness: If the algorithm terminates with a complete and clash-free \CS
  $S$, then the tested formula is satisfiable.
\item Completeness: If the formula is satisfiable, then there is a sequence of
  rule applications that yields a complete and clash-free \CS
\end{enumerate}

The error of the incorrect algorithm is, that is does not satisfy Property 2,
even though the converse is claimed:

\begin{quote}
  \noindent \textsc{Claim}(\cite{JLOGC::HoekR1995}): Let $\phi$ be a
  \grkr-formula in NNF.  $\phi$ is satisfiable iff $\{ x_0 \models \phi \}$ can
  be transformed into a clash-free complete \CS using the rules from
  Figure~\ref{fig:expansion-rules-derijke}.
\end{quote}

Unfortunately, the \emph{if}-direction of this claim is not true, which we will
prove by a simple counterexample. Consider the formula
\[
\phi = \modex R 2 p_1 \wedge \modall R 1 p_2 \wedge \modall R 1 \neg p_2.
\]
On the one hand, $\phi$ is not satisfiable. Assume $\M,x \models \modex R 2
p_1$.  This implies the existence of at least three $R$-successors $y_1,y_2,y_3$
of $x$. For each of the $y_i$
either $\M,y_i \models p_2$ or $\M,y_i \not\models p_2$ holds by the definition
of $\models$. Without loss of generality, there are two worlds $y_{i_1},
y_{i_2}$ such that $\M,y_{i_j} \models p_2$, which implies $\M,x \not\models
\modall R 1 \neg p_2$ and hence  $\M,x \not\models \phi$.

On the other hand, the \CS $S = \{ x \models \phi \}$ can be turned into a
complete and clash-free \CS using the rules from
Fig.~\ref{fig:expansion-rules-derijke}, as is shown in Fig.~\ref{fig:wrong-run}.
Clearly this invalidates the claim and its proof.

\begin{figure}
  \begin{center}
    \begin{align*}
      \{ x \models \phi \} \ruleand \cdots \ruleand & \underbrace{\{ x \models
        \phi, \ x \models \modex R 2 p_1, \ x \models \modall R 1 p_2, \ x
        \models \modall R
        1 \neg p_2 \}}_{=S_1}\\
      \rulegt \cdots \rulegt & \underbrace{S_1 \cup \{ Rxy_i, \ y_i \models p_1
        \mid i = 1, 2, 3 \}}_{= S_2}
    \end{align*}
    $S_2$ is clash-free and complete, because $\sharp R^{S_2}(x,p_1) = 3$ and
    $\sharp R^{S_2}(x,p_2) = 0$.
    \caption{A run of the incorrect algorithm.}
    \label{fig:wrong-run}
  \end{center}
\end{figure}

\subsection{An alternative syntax}

At this stage the reader may have noticed the cumbersome semantics of the
$\modall R n$ operator, which origins from the wish that the duality $\Box \phi
\equiv \neg \Diamond \neg \phi$ of $\mathbf{K}$ carries over to $\modall R n
\phi \equiv \neg \modex R n \neg \phi$ in \grkr. This makes the semantics of
$\modall R n$ and $\modex R n$ un-intuitive. Not only does the $n$ in a diamond
operator mean ``more than $n$'' while it means ``less \emph{or equal} than $n$''
for a box operator. The semantics also introduce a ``hidden'' negation.

To overcome these problems, we will replace these modal operators by a syntax inspired
by the counting quantifiers in predicate logic: the operators $\modleq R n$ and
$\modgeq R n$ with semantics defined by :
\begin{align*}
\M,x \models \modleq R n \phi \ & \text{iff} \ \sharp R^\M(x,\phi) \leq n,\\
\M,x \models \modgeq R n \phi \ & \text{iff} \ \sharp R^\M(x,\phi) \geq n.
\end{align*}
This modification does not change the expressivity of the language, since $\M,x
\models \modex R n \phi$ iff $\M,x \models \modgeq R {n+1} \phi$ and
$\M,x \models \modall R n \phi \ \text{iff} \ \M,x \models \modleq R n \neg
\phi$. We use the following equivalences to transform formulae in the new
syntax into NNF:
\begin{align*}
  \neg \modgeq R 0 \phi & \equiv p \wedge \neg p\\
  \neg \modgeq R n \phi & \equiv \modleq R {n-1} \phi \text{ iff } n >
  1\\
  \neg \modleq R n \phi & \equiv \modgeq R {n+1} \phi
\end{align*}

\subsection{A correct but inefficient solution}

To understand the mistake of the incorrect algorithm, it is useful to know how
soundness is usually established for the kind of algorithms we consider. The
underlying idea is that a complete and clash-free \CS induces a model for the
formula tested for satisfiability:

\begin{definition}[Canonical Structure]
  \label{def:canonical-structure}
  Let $S$ be a c.s. The \emph{canonical structure} $\M_S = (W^{\M_S}, \{R^{\M_S}
  \mid R \in \mathcal{R}\}, V^{\M_S})$ \emph{induced by} $S$ is defined as
  follows:
    \begin{align*}
      W^{\M_S} &= \{ x \in \mathcal{V} \mid \text{$x$ occurs in $S$} \},\\
      R^{\M_S} &= \{ (x,y) \in \mathcal{V}^2 \mid Rxy \in S\},\\
      V^{\M_S}(p) &= \{ x \in \mathcal{V} \mid x \models p \in S\}.
    \end{align*}
\end{definition}

Using this definition, it is then easy to prove that the canonical structure
induced by a complete and clash-free \CS is a model for the tested formula.

The mistake of the incorrect algorithm is due to the fact that it did not take
into account that, in the canonical model induced by a complete and clash-free
c.s., there are formulae satisfied by the worlds even though these formulae do
not appear as constraints in the c.s.  Already in~\cite{HollunderBaader-KR-91},
an algorithm very similar to the incorrect one is presented which decides the
satisfiability of $\alcq$, a notational variant of \grkr.

The algorithm essentially uses the same definitions and rules. The only
differences are the introduction of the $\rulechoose$-rule and an adaption of
the $\rulegeq$-rule to the alternative syntax. The $\rulechoose$-rule makes sure
that all ``relevant'' formulae that are implicitly satisfied by a variable are
made explicit in the c.s. Here, relevant formulae for a variable $y$ are those
occuring in modal formulae in constraints for variables $x$ such that $Rxy$ appears
in the  c.s. The complete rule set for the modified syntax of \grkr is given in
Fig.~\ref{fig:mod-rules}. The definition of \emph{clash} has to be modified as
well: A \CS $S$ contains a clash iff
\begin{itemize}
\item $\{x \models p, x \models \neg p \} \subseteq S$ for some variable $x$ and
  a propositional atom $p$, or
\item $x \models \modleq R n \phi \in S$ and $\sharp R^S(x,\phi) > n$ for some
  variable $x$, relation $R$, formula $\phi$, and $n \in \N$.
\end{itemize}

Furthermore, the notion of safe replacement has to be adapted to the
new syntax: the replacement of $y$ by $z$ in $S$ is called \emph{safe} iff,
for every variable $x$, formula $\phi$, and relation symbol $R$ with $\{x
\models \modgeq R n \phi, Rxy, Rxz\} \subseteq S$ we have $ \sharp
R^{[z/y]S}(x,\phi) \geq n$.

The algorithm, which works like the incorrect algorithm but uses the expansion
rules from Fig.~\ref{fig:mod-rules}---where $\bowtie$ is used as a
placeholder for either $\leq$ or $\geq$---and the definition of clash from above will
be called the \emph{standard algorithm}; it is a decision procedure for
\satgrkr:

\begin{figure}[tbp]
  \begin{center}
    \begin{tabular}{@{ }l@{ }l@{ }l@{}}
      \ruleand-, \ruleor-rule: & see & Fig.~\ref{fig:expansion-rules-derijke}\\[1ex]      
      $\rulechoose$-rule: & if \hfill 1. & $x \models \langle R \rangle_{\bowtie n} \phi, Rxy \in
      S$ and\\
      & \hfill 2. & $\{ y \models \phi, y \models \nneg \phi \} \cap S =
      \emptyset$\\
      & then & $S \rulechoose S \cup \{ y \models \chi \}$ where $\chi \in \{
      \phi, \nneg \phi\}$\\[1ex]
      $\rulegeq$-rule: & if \hfill 1. & $x \models \modgeq R n \phi \in S$ and
      \\
      & \hfill 2. & $\sharp R^S(x,\phi) < n$\\
      & then & $S \rulegeq S \cup \{ R x y, y \models \phi \}$ where $y$ is a new
      variable.  \\[1ex]
      $\ruleleq$-rule: & if \hfill 1. & $x \models \modleq R n \phi, \sharp
      R^S(x,\phi) > n$ and \\
      & \hfill 2. & $y \neq z, Rxy, Rxz, y \models \phi, z \models \phi \in S$
      and \\
      & \hfill 3. & the replacement of $y$ by $z$
      is safe in $S$\\
      & then &$S \ruleleq [y/z]S$
    \end{tabular}
    \caption{The standard completion rules}
    \label{fig:mod-rules}
  \end{center}
\end{figure}

\begin{theorem}[\cite{HollunderBaader-KR-91}]\label{theo:modified-sound-and-correct}
  Let $\phi$ be a \grkr-formula in NNF. $\phi$ is satisfiable iff $\{ x_0
  \models \phi \}$ can be transformed into a clash-free complete \CS using the
  rules in Figure~\ref{fig:mod-rules}. Moreover, each sequence of these
  rule-applications is finite.
\end{theorem}

While no complexity result is explicitly given in \cite{HollunderBaader-KR-91}, it is
easy to see that a \pspace result could be derived from the algorithm using the
trace technique, employed in \cite{Schmidt-Schauss91} to show that
satisfiability of \alc, the notational variant for \kr, is decidable in \pspace.

Unfortunately this is only true if we assume the numbers in the input to be
unary coded. The reason for this lies in the \rulegeq-rule, which generates $n$
successors for a formula of the form $\modgeq R n \phi$. If $n$ is unary coded,
these successors consume at least polynomial space in the size of the input
formula. If we assume binary (or $k$-ary with $k>1$) encoding, the space
consumption is exponential in the size of the input because a number $n$ can be
represented in $\log_k n$ bits in $k$-ary coding. This blow-up can not be
avoided because the completeness of the standard algorithm relies on the
generation \emph{and identification} of these successors, which makes it
necessary to keep them in memory \emph{at one time}.


\section{An optimal solution}

In the following, we will present the algorithm which will be used to prove the
following theorem; it contradicts  the \exptime-hardness conjecture
in~\cite{JLOGC::HoekR1995}.

\begin{theorem}\label{theo:completeness-binary-coding}
  Satisfiability for \grkr is \pspace-complete if numbers in the input are
  represented using \textbf{binary} coding.
\end{theorem}

When aiming for a \pspace algorithm, it is impossible to generate all successors
of a variable in a \CS at a given stage because this may consume space that is
exponential in the size of the input concept. We will give an optimised rule set
for \grkr-satisfiability that does not rely on the identification of successors.
Instead we will make stronger use of non-determinism to guess the assignment
of the relevant formulae to the successors by the time of their generation.
This will make it possible to generate the \CS in a depth first manner, which
will facilitate the re-use of space.

The new set of rules is shown in Fig.~\ref{fig:opt-rules}. The algorithm that
uses these rules is called the \emph{optimised algorithm}.  The definition of
\emph{clash} is taken from the standard algorithm. We do not need a
\ruleleq-rule.

\begin{figure}[b]
  \begin{center}
    \begin{tabular}{@{ }l@{ }l@{ }l@{}}
      \multicolumn{3}{@{ }l}{$\ruleand$-, $\ruleor$-rule: see Fig.~\ref{fig:expansion-rules-derijke}}\\[1ex]      
      $\rulegeq$-rule: & if \hfill 1. & $x \models \modgeq R n \phi \in S$, and
      \\
      & \hfill 2. & $\sharp R^S(x,\phi) < n$, and \\
      & \hfill 3. & neither the $\ruleand$- nor the $\ruleor$-rule apply to a constraint
      for $x$ \\
      & then & $ S \rulegeq S \cup \{ Rxy, y \models \phi, y \models \chi_1, \dots, y
      \models \chi_k \}$  where \\
      & & $\{ \psi_1, \dots, \psi_k \} = \{ \psi \mid x \models \langle R
      \rangle_{\bowtie m} \psi \in S \}$, $\chi_i \in \{ \psi_i, \nneg \psi_i
      \}$, and\\
      & & $y$ is a fresh variable.
    \end{tabular}
  \end{center}
  \caption{The optimised completion rules.}
  \label{fig:opt-rules}
\end{figure}

At first glance, the $\rulegeq$-rule may appear to be complicated and therefor
is explained in more detail: like the standard $\rulegeq$-rule, it is
applicable to a \CS that contains the constraint $x \models \modgeq R n \phi$
if there are less than $n$ $R$-successors $y$ of $x$ with $y \models \phi \in
S$.  The rule then adds a new successor $y$ to $S$.  Unlike the standard
algorithm, the optimised algorithm also adds additional constraints of the
form $y \models (\nneg)\psi$ to $S$ for each formula $\psi$ appearing in a
constraint of the form $x \models \langle R \rangle _{\bowtie n} \psi$. Since
we have suspended the application of the \rulegeq-rule until no other rule
applies to $x$, by this time $S$ contains all constraints of the form $x
\models \langle R \rangle_{\bowtie n} \psi$ it will ever contain. This
combines the effects of both the $\rulechoose$- and the $\ruleleq$-rule of the
standard algorithm.

\subsection{Correctness of the optimised algorithm}

To establish the correctness of the optimised algorithm, we will show its
termination, soundness, and completeness.

To analyse the memory usage of the algorithm it is very helpful to view a \CS
as a graph: A \CS $S$ induces a labeled graph $G(S) = (N,E,\mathcal{L})$ with
\begin{itemize}
\item The set of nodes $N$ is the set of variables appearing in $S$.
\item The edges $E$ are defined by $E := \{ xy \mid R x y \in S \ \text{for some
    $R \in \mathcal{R}$} \}$.
\item $\mathcal{L}$ labels nodes and edges in the following way:
  \begin{itemize}
  \item For a node $x \in N$: $\mathcal{L}(x) := \{ \phi \mid x \models \phi
    \in S \}$.
  \item For an edge $xy \in E$: $\mathcal{L}(xy) := \{ R \mid Rxy \in S\}$.
  \end{itemize}
\end{itemize}

It is easy to show that the graph $G(S)$ for a \CS $S$ generated by the
optimised algorithm from an initial \CS $\{ x_0 \models \phi \}$ is a tree with
root $x_0$, and for each edge $xy \in E$, the label $\mathcal{L}(xy)$ is a
singleton.  Moreover, for each $x \in N$ it holds that $\mathcal{L}(x) \subseteq
\sub(\phi)$ where $\sub(\phi)$ is the smallest set of formulae satisfying
\begin{itemize}
\item $\phi \in \sub(\phi)$,
\item if $\psi_1 \vee \psi_2 \ \text{or} \ \psi_1 \wedge \psi_2 \in \sub(\phi)$,
  then also $\psi_1,\psi_2 \in \sub(\phi)$,
\item if $\langle R \rangle_{\bowtie n} \psi \in \sub(\phi)$, then also
  $\psi \in \sub(\phi)$,
\item if $\psi \in \sub(\phi)$, then also $\nneg \psi \in \sub(\phi)$.
\end{itemize}

We will use the fact that the number of elements of $\sub(\phi)$ is bounded by
$2 \times |\phi|$ where $|\phi|$ denotes the length of $\phi$. This is easily
shown by proving
\[
\sub(\phi) = \textit{sub}(\phi) \cup \{ \nneg \psi \mid \psi \in
\textit{sub}(\phi) \}
\]
where $\textit{sub}(\phi)$ denotes the set of all sub-formulae of $\phi$. The
size of $\textit{sub}(\phi)$ is obviously bounded by $|\phi|$.

\subsubsection{Termination}

First, we will show that the optimised algorithm always terminates, i.e., each
sequence of rule applications starting from a \CS of the form $\{ x_0 \models
\phi\}$ is finite. The next lemma will also be of use when we will consider the
complexity of the algorithm.

\begin{lemma}\label{lem:path-is-poly-long}
  Let $\phi$ be a formula in NNF and $S$ a \CS that is generated by the
  optimised algorithm starting from $\{ x_0 \models \phi \}$.
  \begin{itemize}
  \item The length of a path in $G(S)$ is limited by $|\phi|$.
  \item The out-degree of $G(S)$ is bounded by $|\sub(\phi)| \times 2^{|\phi|}$.
  \end{itemize}
\end{lemma}

\begin{proof}
  For a variable $x \in N$, we define $\ell(x)$ as the maximum depth of nested
  modal operators in $\mathcal{L}(x)$. Obviously, $\ell(x_0) \leq |\phi|$ holds. Also,
  if $xy \in E$ then $\ell(x) > \ell(y)$. Hence each path $x_1,\dots,x_k$ in
  $G(S)$ induces a sequence $\ell(x_1) > \dots > \ell(x_k)$ of natural numbers.
  $G(S)$ is a tree with root $x_0$, hence the longest path in $G(S)$ starts with
  $x_0$ and its length is bounded by $|\phi|$.
  
  Successors in $G(S)$ are only generated by the $\rulegeq$-rule. For a variable
  $x$ this rule will generate at most $n$ successors for each $\modgeq R n \psi
  \in \mathcal{L}(x)$. There are at most $|\sub(\phi)|$ such formulae in
  $\mathcal{L}(x)$. Hence the out-degree of $x$ is bounded by $|\sub(\phi)|
  \times 2^{|\phi|}$, where $2^{|\phi|}$ is a limit for the biggest number that
  may appear in $\phi$ if binary coding is used.  \qed
\end{proof}

\begin{corollary}[Termination]
\label{lem:optimised-alg-terminates}
Any sequence of rule applications starting from a \CS $S = \{ x_0 \models \phi
\}$ of the optimised algorithm is finite.
\end{corollary}

\begin{proof}
  The sequence of rules induces a sequence of trees. The depth and the
  out-degree of these trees is bounded in $|\phi|$ by
  Lemma~\ref{lem:path-is-poly-long}. For each variable $x$ the label
  $\mathcal{L}(x)$ is a subset of the finite set $\sub(\phi)$. Each application
  of a rule either
  \begin{itemize}
  \item adds a constraint of the form $x \models \psi$ and hence adds an element
    to $\mathcal{L}(x)$, or
  \item adds fresh variables to $S$ and hence adds additional nodes to the
    tree $G(S)$.
  \end{itemize}
  Since constraints are never deleted and variables are never identified, an
  infinite sequence of rule application must either lead to an arbitrary large
  number of nodes in the trees which contradicts their boundedness, or it leads
  to an infinite label
  of one of the nodes $x$ which contradicts $\mathcal{L}(x) \subseteq
  \sub(\phi)$.
  \qed
\end{proof}

\subsubsection{Soundness and Completeness}

The following definition will be very helpful to establish soundness and
completeness of the optimised algorithm:

\begin{definition}\label{def:constraint-system-satisfiable}
  A \CS $S$ is called \emph{satisfiable} iff there exists a Kripke structure
  $\M = (W^\M, \{R^\M \mid R\in \mathcal{R}\}, V^\M)$ and a mapping $\alpha :
  \mathcal{V} \rightarrow W^\M$ such that the following properties hold:
  \begin{enumerate}
  \item If $y,z$ are distinct variables such that $Rxy,Rxz \in S$, then $\alpha(y) \neq
    \alpha(z)$.
  \item If $x \models \psi \in S$ then $\M,\alpha(x) \models \psi$.
  \item If $Rxy \in S$ then $(\alpha(x),\alpha(y)) \in R^\M$.
  \end{enumerate}
  In this case, $\M,\alpha$ is called a \emph{model} of $S$.
\end{definition}

It easily follows from this definition, that a \CS $S$ that contains a clash
can not be satisfiable and that the \CS  $\{ x_0 \models \phi \}$ is
satisfiable if and only if $\phi$ is satisfiable.

\begin{lemma}[Local Correctness]
  \label{lem:local-soundness}
  Let $S,S'$ be \CS generated by the optimised algorithm from a
  \CS of the form $\{ x_0 \models \phi \}$.
  \begin{enumerate}
  \item If $S'$ is obtained from $S$ by application of the (deterministic)
    $\ruleand$-rule, then $S$ is satisfiable if and only if $S'$ is satisfiable.
  \item If $S'$ is obtained from $S$ by application of the (non-deterministic)
    $\ruleor$- or $\rulegeq$-rule, then $S$ is satisfiable if $S'$ is
    satisfiable. Moreover, if $S$ is satisfiable, then the rule can always be
    applied in such a way that it yields a \CS $S'$ that is satisfiable.
  \end{enumerate}
\end{lemma}

\begin{proof}
  $S \rightarrow S'$ for any rule $\rightarrow$ implies $S \subseteq S'$, hence
  each model of $S'$ is also a model of $S$.  Consequently, we must show only
  the other direction. 
  \begin{enumerate}
  \item Let $\M,\alpha$ be a model of $S$ and let $x \models \psi_1 \wedge
    \psi_2$ be the constraint that triggers the application of the
    $\ruleand$-rule. The constraint $x \models \psi_1 \wedge \psi_2 \in S$
    implies $\M,\alpha(x) \models \psi_1 \wedge \psi_2$.  This implies
    $\M,\alpha(x) \models \psi_i$ for $i=1,2$. Hence $\M,\alpha$ is also a model
    of $S' = S \cup \{ x \models \psi_1, x \models \psi_2 \}$.
  \item Firstly, we consider the $\ruleor$-rule. Let $\M,\alpha$ be a model of
    $S$ and let $x \models \psi_1 \vee \psi_2$ be the constraint that triggers
    the application of the $\ruleor$-rule. $x \models \psi_1 \vee \psi_2 \in S$
    implies $\M,\alpha(x) \models \psi_1 \vee \psi_2$.  This implies
    $\M,\alpha(x) \models \psi_1$ or $\M,\alpha(x) \models \psi_2$.  Without
    loss of generality we may assume $\M,\alpha(x) \models \psi_1$. The
    $\ruleor$-rule may choose $\chi = \psi_1$, which implies $S' = S \cup \{ x
    \models \psi_1 \}$ and hence $\M,\alpha$ is a model for $S'$.
    
    Secondly, we consider the $\rulegeq$-rule. Again let $\M,\alpha$ be a model
    of $S$ and let $x \models \modgeq R n \phi$ be the constraint that triggers
    the application of the $\rulegeq$-rule. Since the $\rulegeq$-rule is
    applicable, we have $\sharp R^S(x,\phi) < n$. We claim that there is a $w
    \in W^\M$ with
    \begin{equation}
      \tag{$*$}
      (\alpha(x),w) \in R^\M, \M,w \models \phi, \ \text{and} \ w \not\in \{ \alpha(y) \mid
      Rxy \in S \}.
    \end{equation}
    Before we prove this claim, we show how it can be used to finish the proof.
    The world $w$ is used to ``select'' a choice of the \rulegeq-rule that
    preserves satisfiability: Let $\{ \psi_1, \dots, \psi_n \}$ be an
    enumeration of the set $\{ \psi \mid x \models \langle R \rangle_{\bowtie n}
    \psi \in S \}$. We set
    \[
    S' = S \cup \{Rxy, y \models \phi\} \cup \{ y \models \psi_i \mid \M,w
    \models \psi_i \} \cup \{ y \models \nneg \psi_i \mid \M,w \not\models
    \psi_i \}.
    \]
    Obviously, $\M,\alpha[y \mapsto w]$ is a model for $S'$ (since $y$ is a
    fresh variable and $w$ satisfies $(*)$), and $S'$ is a possible result of
    the application of the $\rulegeq$-rule to $S$.
  \end{enumerate}
  
  We will now come back to the claim. It is obvious that there is a $w$ with
  $(\alpha(x),w) \in R^\M$ and $\M,w \models \phi$ that is not contained in $\{
  \alpha(y) \mid Rxy, y \models \phi \in S \}$, because $\sharp R^\M(x,\phi) \geq
  n > \sharp R^S(x,\phi)$.  Yet $w$ might appear as the image of an element $y'$
  such that $Rxy' \in S$ but $y' \models \phi \not\in S$.  
  
  Now, $Rxy' \in S$ and $y' \models \phi \not\in S$ implies $y' \models
  \nneg\phi \in S$.  This is due to the fact that the constraint $Rxy'$ must
  have been generated by an application of the $\rulegeq$-rule because it has
  not been an element of the initial c.s. The application of this rule was
  suspended until neither the $\ruleand$- nor the $\ruleor$-rule are applicable
  to $x$.  Hence, if $x \models \modgeq R n \phi$ is an element of $S$ now,
  then it has already been in $S$ when the $\rulegeq$-rule that generated $y'$
  was applied. The $\rulegeq$-rule guarantees that either $y' \models \phi$ or
  $y' \models \nneg \phi$ is added to $S$. Hence $y' \models \nneg \phi \in S$.
  This is a contradiction to $\alpha(y') = w$ because under the assumption that
  $\M,\alpha$ is a model of $S$ this would imply $\M,w \models \nneg \phi$ while
  we initially assumed $\M,w \models \phi$. \qed
\end{proof}

From the local completeness of the algorithm we can immediately derive the
global completeness of the algorithm:

\begin{lemma}[Completeness]
  \label{lem:completeness-optimised-algorithm}
  If $\phi \in \satgrkr$ in NNF, then there is a sequence of applications of the
  optimised rules starting with $S = \{ x_0 \models \phi\}$ that results in a
  complete and clash-free c.s.
\end{lemma}

\begin{proof}
  The satisfiability of $\phi$ implies that also $\{ x_0 \models \phi \}$ is
  satisfiable. By Lemma~\ref{lem:local-soundness} there is a sequence of
  applications of the optimised rules which preserves the satisfiability of the
  c.s. By Lemma~\ref{lem:optimised-alg-terminates} any sequence of applications
  must be finite. No generated \CS (including the last one) may contain a clash
  because this would make it unsatisfiable. \qed
\end{proof}

Note that since we have made no assumption about the order in which the rules
are applied (with the exception that is stated in the conditions of the
$\rulegeq$-rule), the selection of the constraints to apply a rule to as well as
the selection which rule to apply is ``don't-care'' non-deterministic, i.e., if a
formula is satisfiable, then this can be proved by an arbitrary sequence of rule
applications. Without this property, the resulting algorithm certainly would be
useless for practical applications, because any deterministic implementation
would have to use backtracking for the selection of constraints and
rules.

\begin{lemma}[Soundness]
  \label{lem:soundness-optimised-algorithm}
  Let $\phi$ be a \grkr-formula in NNF. If there is a sequence of applications
  of the optimised rules starting with the \CS $\{ x_0 \models
  \phi \}$ that results in a complete and clash-free c.s., then
  $\phi \in \satgrkr$.
\end{lemma}

\begin{proof}
  Let $S$ be a complete and clash-free \CS generated by applications of the
  optimised rules. We will show that the canonical model $\M_S$ together with
  the identity function is a model for $S$. Since $S$ was generated from $\{ x_0
  \models \phi\}$ and the rules do not remove constraints from the c.s., $x_0
  \models \phi \in S$. Thus $\M_S$ is also a model for $\phi$ with $\M_S,x_0
  \models \phi$.

  By construction of $\M_S$, Property~1 and 3 of
  Definition~\ref{def:constraint-system-satisfiable} are trivially satisfied.
  It remains to show that $x \models \psi \in S$ implies $\M_S,x \models \psi$,
  which we will show by induction on the norm $\|\cdot\|$ of a formula $\psi$.
  The norm $\| \psi \|$ for formulae in NNF is inductively defined by:
  \[
  \begin{array}{lclcl}
    \|p\| & := & \|\neg p\| & := & 0 \quad \text{for $p\in\mathcal{P}$}\\ 
    \|\psi_1 \wedge \psi_2 \| & := &  \| \psi_1 \vee \psi_2\| & := & 
    1+\|\psi_1\|+\|\psi_2\|\\
    \|\langle R \rangle_{\bowtie n} \psi\| & & & := & 1+\|\psi\|
  \end{array} 
  \]
  This definition is chosen such that it satisfies $\|\psi\| = \| \nneg \psi \|$
  for every formula $\psi$.
  \begin{itemize}
  \item The first base case is $\psi = p$ for $p \in \mathcal{P}$. $x \models p \in
    S$ implies $x \in V^{\M_S}(p)$ and hence ${\M_S},x \models p$.  The second base
    case is $x \models \neg p \in S$. Since $S$ is clash-free, this implies $x
    \models p \not\in S$ and hence $x \not\in V^{\M_S}(p)$. This implies
    ${\M_S},x \models \neg p$.
  \item $x \models \psi_1 \wedge \psi_2 \in S$ implies $x \models \psi_1, x
    \models \psi_2 \in S$. By induction, we have ${\M_S},x \models \psi_1$ and
    ${\M_S},x \models \psi_2$ holds and hence ${\M_S},x \models \psi_1 \wedge
    \psi_2$.  The case $x \models \psi_1 \vee \psi_2 \in S$ can be handled
    analogously.
  \item $x \models \modgeq R n \psi \in S$ implies $\sharp R^S(x,\psi) \geq n$
    because otherwise the $\rulegeq$-rule would be applicable and $S$ would not
    be complete. By induction, we have ${\M_S},y \models \psi$ for each $y$ with
    $y \models \psi \in S$. Hence $\sharp R^{\M_S}(x,\psi) \geq n$ and thus
    ${\M_S},x \models \modgeq R n \psi$.
  \item $x \models \modleq R n \psi \in S$ implies $\sharp R^S(x,\psi) \leq n$
    because $S$ is clash-free.  Hence it is sufficient to show that $\sharp
    R^{\M_S}(x,\psi) \leq \sharp R^S(x,\psi)$ holds. On the contrary, assume
    $\sharp R^{\M_S}(x,\psi) > \sharp R^S(x,\psi)$ holds. Then there is a
    variable $y$ such that $Rxy \in S$ and ${\M_S},y \models \psi$ while $y
    \models \psi \not\in S$.  For each variable $y$ with $Rxy \in S$ either $y
    \models \psi \in S$ or $y \models \nneg \psi \in S$.  This implies $y
    \models \nneg \psi \in S$ and, by the induction hypothesis, ${\M_S},y \models
    \nneg \psi$ holds which is a contradiction. \qed
  \end{itemize} 
\end{proof}

The following theorem is an immediate consequence of
Lemma~\ref{lem:optimised-alg-terminates},
\ref{lem:completeness-optimised-algorithm}, and
\ref{lem:soundness-optimised-algorithm}:

\begin{corollary}\label{theo:correctness-optimised-algorithm}
  The optimised algorithm is a non-deterministic decision procedure for
  \satgrkr.
\end{corollary}

\subsection{Complexity of the optimised algorithm}

The optimised algorithm will enable us to prove
Theorem~\ref{theo:completeness-binary-coding}. We will give a proof by sketching
an implementation of this algorithm that runs in polynomial space. 

\begin{lemma}\label{lem:optimised-algorithm-in-pspace}
  The optimised algorithm can be implemented in \pspace
\end{lemma}

\begin{proof}
  Let $\phi$ be the \grkr-formula to be tested for satisfiability. We can assume
  $\phi$ to be in NNF because the transformation of a formula to NNF can be
  performed in linear time and space.
  
  The key idea for the \pspace implementation is the \emph{trace
    technique}~\cite{Schmidt-Schauss91}, i.e., it is sufficient to keep only a
  single path (a trace) of $G(S)$ in memory at a given stage if the \CS is
  generated in a depth-first manner. This has already been the key to a \pspace
  upper bound for \kr and \alc
  in~\cite{SICOMP::Ladner1977,Schmidt-Schauss91,HalpernMoses92}. To do this we
  need to store the values for $\sharp R^S(x,\psi)$ for each variable $x$ in the
  path, each $R$ which appears in $\sub(\phi)$ and each $\psi \in \sub(\phi)$.
  By storing these values in binary form, we are able to keep information
  \emph{about} exponentially many successors in memory while storing only a
  single path at a given stage.
  
  Consider the algorithm in Fig.~\ref{fig:decision-proc}, where
  $\mathcal{R}_\phi$ denotes the set of relation names that appear in $\sub(\phi)$.
  It re-uses the space needed to check the satisfiability of a successor $y$ of
  $x$ once the existence of a complete and clash-free ``subtree'' for the
  constraints on $y$ has been established.  This is admissible since the
  optimised rules will never modify this subtree once is it completed.
  Neither do constraints in this subtree have an influence on the completeness
  or the existence of a clash in the rest of the tree, with the exception that
  constraints of the form $y \models \psi$ for $R$-successors $y$ of $x$
  contribute to the value of $\sharp R^S(x,\psi)$. These numbers play a role
  both in the definition of a clash and for the applicability of the
  $\rulegeq$-rule. Hence, in order to re-use the space occupied by the subtree
  for $y$, it is necessary and sufficient to store these numbers.

  \begin{figure}[t]
    \begin{center}
      \leavevmode
      \begin{tabbing}
        \hspace{2em}\=\hspace{2em}\=\hspace{2em}\= \kill
        $\grkr-\textsc{SAT}(\phi) := \texttt{sat}(x_0,\{x_0 \models \phi\})$\\
        $\texttt{sat}(x,S)$: \\
        \> allocate counters  $\sharp R^S(x,\psi) :=0$ for all $R \in
        \mathcal{R}_\phi$ and $\psi \in \sub(\phi)$. \\
        \> \texttt{while} (the $\ruleand$- or the $\ruleor$-rule can be
        applied) \texttt{and}  ($S$ is clash-free) \texttt{do}\\
        \>\> apply the $\ruleand$- or the $\ruleor$-rule to $S$.\\
        \> \texttt{od}\\
        \> \texttt{if} $S$ contains a clash \texttt{then}
        \texttt{return} ``not satisfiable''.\\
        \> \texttt{while} (the $\rulegeq$-rule applies to $x$ in $S$) \texttt{do} \\
        \>\> $S_{\textit{new}} := \{Rxy, y \models \phi', y \models \chi_1, \dots, y
        \models \chi_k \}$ \\
        \>\> \texttt{where}\\
        \>\>\> $y$ is a fresh variable,\\
        \>\>\> $x \models \modgeq R n \phi'$ triggers an application of the
        $\rulegeq$-rule,\\
        \>\>\> $\{ \psi_1, \dots, \psi_k \} = \{ \psi \mid x \models \langle R
        \rangle_{\bowtie  n} \psi \in S \}$, and \\
        \>\>\> $\chi_i$ is chosen non-deterministically from $\{ \psi_i, \nneg \psi_i \}$ \\
        \>\> \texttt{for each} $y \models \psi \in S_{\textit{new}}$ \texttt{do}
        increment $\sharp R^S(x,\psi)$\\
        \>\> \texttt{if} $x \models \modleq R m \psi \in S$ and $\sharp
        R^S(x,\psi) > m$ \texttt{then}
        \texttt{return} ``not satisfiable''.\\
        \>\> \texttt{if} $\texttt{sat}(y,S_{\textit{new}}) = 
        \text{``not satisfiable''}$ \texttt{then} \texttt{return} ``not satisfiable'' \\[0.25ex]
        \> \texttt{od}\\
        \> remove the counters for $x$ from memory.\\
        \> \texttt{return} ``satisfiable''
      \end{tabbing}
      \caption{A non-deterministic \pspace decision procedure for \satgrkr.}
      \label{fig:decision-proc}
    \end{center}
  \end{figure}
  
  Let us examine the space usage of this algorithm. Let $n = |\phi|$. The
  algorithm is designed to keep only a single path of $G(S)$ in memory at a
  given stage.  For each variable $x$ on a path, constraints of the form $x
  \models \psi$ have to be stored for formulae $\psi \in \sub(\phi)$. The size
  of $\sub(\phi)$ is bounded by $2n$ and hence the constraints for a single
  variable can be stored in $\mathcal{O}(n)$ bits.  For each variable, there are
  at most $|\mathcal{R}_\phi| \times |\sub(\phi)| = \mathcal{O}(n^2)$ counters
  to be stored. The numbers to be stored in these counters do not exceed the
  out-degree of $x$, which, by Lemma~\ref{lem:path-is-poly-long}, is bounded by
  $|\sub(\phi)|\times 2^{|\phi|}$. Hence each counter can be stored using
  $\mathcal{O}(n^2)$ bits when binary coding is used to represent the counters,
  and all counters for a single variable require $\mathcal{O}(n^4)$ bits.  Due
  to Lemma~\ref{lem:path-is-poly-long}, the length of a path is limited by $n$,
  which yields an overall memory consumption of $\mathcal{O}(n^5 + n^2)$. \qed
\end{proof}

Theorem~\ref{theo:completeness-binary-coding} now is a simple Corollary from the
\pspace-hardness of \kr, Lemma~\ref{lem:optimised-algorithm-in-pspace}, and
Savitch's Theorem \cite{JCSS::Savitch1970}.



\section{Extensions of the Language}

It is possible to extend the language \grkr without loosing the \pspace
property of the satisfiability problem. In this section we extend the
techniques to obtain a \pspace algorithm for the logic \grkrri, which extends
\grkr by intersection of accessibility relations and inverse relations. These
extension are mainly motivated from the world of Description Logics, where they
are commonly studied. In this context, the logic \grkrri can be perceived as a
notational variant of the Description Logic $\ALCQIR$.

\begin{definition}[Syntax and Semantics of \grkrri]
  \label{def:grknri-syntax+semantics}
  Let $\mathcal{P} = \{p_0, p_1, \dots \}$ be a set of proposition letters and let
  $\mathcal{R}$ be a set of \emph{relation names}.  The set $\overline
  {\mathcal{R}} := \mathcal{R} \cup \{ R^{-1} | R \in \mathcal{R} \}$ is called
  the set of \grkrri-relations.

  The set of
  \grkrri-\emph{formulae} is the smallest set such that
  \begin{enumerate}
  \item every proposition letter is a \grkrri-formula and,
  \item if $\phi,\psi_1,\psi_2$ are formulae, $n \in \N$, and $R_1,\dots,R_k$
    are (possibly inverse) \grkrri-relations, then $\neg \phi$, $\psi_1 \wedge
    \psi_2$, $\psi_1 \vee \psi_2$, $\modleq {R_1 \cap \dots \cap R_k} n
    \phi$, and $\modgeq {R_1 \cap \dots \cap R_k} n \phi$ are
    \grkrri-formulae.
  \end{enumerate}

  The semantics are extended accordingly:
  \begin{align*}
    \M,x & \models \modleq {R_1 \cap \dots \cap R_k} n \phi \ \text{iff} \
    \sharp (R_1 \cap \dots \cap R_k)^\M(x,\phi) \leq n\\
    \M,x & \models \modgeq {R_1 \cap \dots \cap R_k} n \phi \ \text{iff} \
    \sharp (R_1 \cap \dots \cap R_k)^\M(x,\phi) \geq n
  \end{align*}
  where
  \[
  \sharp (R_1 \cap \dots \cap R_k)^\M(x,\phi) =  | \{ y \in W^\M \mid (x,y) \in
  R_1^\M \cap \dots \cap R_k^\M  \ \text{and} \ \M,y \models \phi \}|,
  \]
  and, for $R \in \mathcal{R}$, we define
  \[
  (R^{-1})^\M := \{ (y,x) \mid (x,y) \in R^\M \} .
  \]
\end{definition}

We will use the letters $\omega,\sigma$ to range over intersections of
\grkrri-relations. By abuse of notation we will sometimes identify an
intersection of relations $\omega$ with the set of relations occurring in it
and write $R \in \omega$ iff $\omega = R_1 \cap \dots \cap R_k$ and there is
some $1\leq i \leq k$ with $R = R_i$. To avoid dealing with relations of the
form $(R^{-1})^{-1}$ we use the convention that $(R^{-1})^{-1} = R$ for
any $R \in \mathcal{R}$.

Obviously every \grkr formula is also a
\grkrri formula.  Using standard bisimluation arguments one can show that
\grkrri is strictly more expressive than \grkr.

\subsection{Reasoning for \grkrri}

We will use similar techniques as in the previous section to obtain a
\pspace-algorithm for \grkrri. The definition of a constraint system remains
unchanged, but we additionally require that, for any $R \in \mathcal{R}$, a
\CS $S$ contains the constraint `$Rxy$' iff it contains the constraint
`$R^{-1}yx$'. For a \CS $S$, an intersection of \grkrri-relations $\omega =
R_1 \cap \dots \cap R_k$, and a formula $\phi$, let $\sharp \omega^S(x,\phi)$
be the number of variables $y$ such that $\{R_1 x y,\dots, R_k x y, y \models
\phi\} \subseteq S$.

We modify the definition of \emph{clash} to deal with intersection of relations
as follows. A \CS $S$ contains a clash iff
\begin{itemize}
\item $\{x \models p, x \models \neg p \} \subseteq S$ for some variable $x$ and
  a proposition letter $p$, or
\item $x \models \modleq \omega n \phi \in S$ and $\sharp
  \omega^S(x,\phi) > n$ for some variable $x$, intersection of
  \grkrri-relations $\omega$, formula $\phi$ and $n \in \N$.
\end{itemize}

The set of rules dealing with the extended logic is shown in
Figure~\ref{fig:expansion-rules-grkrri}. We require the algorithm to maintain
a binary relation $\prec_S$ between the variables in a \CS $S$ with $x \prec_S
y$ iff $y$ was inserted by the $\rulegeq$-rule to satisfy a constraint for
$x$. When considering the graph $G(S)$, the relation $\prec_S$ corresponds to
the successor relation between nodes. Hence, when $x \prec_S y$ holds we will
call $y$ a successor of $x$ and $x$ a predecessor of $y$. We denote the
transitive closure of $\prec_S$ by $\prec^+_S$. For a set of variables
$\mathcal{X}$ and a \CS $S$, we denote the subset of $S$ in which no variable
from $\mathcal{X}$ occurs in a constraint by $S - \mathcal{X}$.  The
$\ruleand$-, $\ruleor$- and $\rulechoose$-rule are called ``non-generating
rules'' while the $\rulegeq$-rule is called a ``generating rule''. The algorithm
which uses these rules will be called the \grkrri-algorithm.

\begin{figure}
  \begin{center}
    \begin{tabular}{@{ }l@{ }l@{ }l@{}}
      \multicolumn{3}{@{ }l}{\ruleand-, \ruleor-rule:  see 
      Fig.~\ref{fig:expansion-rules-derijke}}\\[1ex]      
      $\rulechoose$-rule: & if \hfill 1. & $x \models \langle \omega
      \rangle_{\bowtie n} \phi \in S$ and\\
      & \hfill 2. & for some $R \in \omega$ there is a $y$ with  $Rxy \in S$, and \\
      && $\{ y \models \phi, y \models \nneg \phi \} \cap S = \emptyset$\\
      & then & $S \rulechoose S' \cup \{ y \models \chi \}$ where $\chi \in \{ \phi,
      \nneg \phi \}$\\
      && and $S' = S - \{ z \mid y \prec^+_S z \}$\\

      $\rulegeq$-rule: & if \hfill 1. & $x \models \modgeq \omega n \phi \in S$, and
      \\
      & \hfill 2. & $\sharp \omega^S(x,\phi) < n$, and \\
      & \hfill 3. & no non-generating rule can be applied to a constraint
      for $x$ \\
      & then & $ S \rulegeq S \cup \{ y \models \psi \} \cup S' \cup S''$ and set $x \prec_S y$ where\\
      & & $S' = \{ y \models \chi_1, \dots, y \models \chi_k \}$, $\chi_i \in \{ \psi_i, \nneg
      \psi_i \}$, and\\
      & & \quad $\{ \psi_1, \dots, \psi_k \} = \{ \psi \mid x \models \langle \sigma
      \rangle_{\bowtie m} \psi \in S \}$\\
      & & $S'' = \{ R_1xy, R^{-1}_1yx, \dots, R_mxy, R^{-1}_myx \}$ and\\
      & & \quad $\omega \subseteq \{ R_1, \dots, R_m \} \subseteq
      \overline{\mathcal{R}}$\\
      && $y$ is a fresh variable\\
    \end{tabular}
  \end{center}
  \caption{The completion rules for \grkrri.}
  \label{fig:expansion-rules-grkrri}
\end{figure}

The $\rulegeq$-rule, while looking complicated, is a straightforward extension
of the $\rulegeq$-rule for \grkr, which takes into account that we also need
to guess additional \emph{relations} between the old variable $x$ and the freshly
introduced variable $y$. The $\rulechoose$-rule requires more explanation.

For \grkr, the optimised algorithm generates a \CS $S$ in a way that, whenever
$x \models \langle R \rangle_{\bowtie n} \psi \in S$, then, for any $y$ with
$Rxy \in S$, either $y \models \psi \in S$ of $y \models \nneg \psi \in S$.
This was achieved by suspending the generation of any successors $y$ of $x$
until $S$ contained all constraints of the from $x \models \phi$ it would ever
contain. In the presence of inverse relations, this is no longer possible
because $y$ might be generated as a predecessor of $x$ and hence before it was
possible to know which $\psi$ might be relevant.  There are at least two
possible ways to overcome this problem. One is, to guess, for every $x$ and
\emph{every} $\psi \in \sub(\phi)$, whether $x \models \psi$ or $x \models
\nneg \psi$. In this case, since the termination of the optimised algorithm as
shown in Lemma~\ref{lem:optimised-alg-terminates} relies on the fact that the
modal depth strictly decreases along a path in the induced graph $G(S)$,
termination would no longer be guaranteed. It would have to be enforced by
different means.

Here, we use another approach. We can distinguish two different situations
where $\{x \models \langle \omega \rangle_{\bowtie n} \psi, Rxy \} \subseteq
S$ for some $R \in \omega$, and $ \{ y \models \psi, y \models \nneg
\psi \} \cap S = \emptyset$, namely, whether $y$ is a predecessor of $x$  ($y
\prec_S x$) or a successor of $x$ ($x \prec_S y$).  The second situation will never
occur. This is due to the interplay of the $\rulegeq$-rule, which is suspended
until all known relevant information has been added for $x$, and the
$\rulechoose$-rule, which deletes certain parts of the \CS whenever new
constraints have to be added for predecessor variables.

The first situation is resolved by non-deterministically adding either $y
\models \psi$ or $y \models \nneg \psi$ to $S$. The subsequent deletion of all
constraints involving variables from $\{ z \mid y \prec^+_S z \}$, which
corresponds to all subtrees of $G(S)$ rooted at successors of $y$, is
necessary to make this rule ``compatible'' with the trace-technique we want to
employ in order to obtain a \pspace-algorithm. The correctness of the
trace-approach relies on the property that, once we have established the
existence of a complete and clash-free ``subtree'' for a node $x$, we can
remove this tree from memory because it will not be modified by the algorithm.
In the presence of inverse relations this can be no longer taken for granted
as can be shown by the formula
\[
\phi = \modleq {R_1} 0 q \; \wedge \; \modgeq {R_1} 1 (p \vee q) \; \wedge \; \modgeq
{R_2} 1 \modleq {R_2^{-1}} 0 \modgeq {R_1} 1 p
\]
Figure~\ref{fig:reset-restart} shows the beginning of a run of the algorithm
for \grkrri. After a number of steps, a successor $y$ of $x$ has been
generated and the expansion of constraints has produced a complete and
clash-free subtree for $y$. Nevertheless, the formula $\phi$ is not
satisfiable. The expansion of $\modgeq {R_2} 1 \modleq {R_2^{-1}} 0 \modgeq
{R_1} 1 p$ will eventually lead to the generation of the constraint $x \models
\nneg \modgeq {R_1} 1 p = \modleq {R_1} 0 p$, which clashes with $y \models
p$. If the subtree for $y$ would already have been deleted from memory, this
clash would go undetected. For this reason, the $\rulechoose$-rule deletes all 
successors of the modified node, which, while duplicating some work, makes it
possible to detect these clashes even when tracing through the c.s. A similar
technique has been used in \cite{HoSatTo-LPAR-99} to obtain a \pspace-result
for a Description Logic with inverse roles.

\begin{figure}
  \begin{center}
    \begin{align*}
      \{ x \models \phi \}& \ruleand \dots \\
      & \ruleand \underbrace{\{ x \models \phi, x \models
        \modleq {R_1} 0 q, x \models \modgeq {R_1} 1 (p \vee q), x \models
        \modgeq {R_2} 1 \modleq {R_2^{-1}} 0 \modgeq {R_1} 1 p \}}_{S_1}\\
      & \rulegeq \underbrace{S_1 \cup \{R_1xy, R^{-1}_1yx, y \models (p \vee q), y \models
        \neg q \}}_{S_2} \ruleor \underbrace{S_2 \cup \{y \models p \}}_{S_3}
    \end{align*}
  \end{center}
  \caption{Inverse roles make tracing difficult.}
  \label{fig:reset-restart}
\end{figure}

\subsection{Correctness of the Algorithm}

As for \grkr, we have to show termination, soundness, and correctness of the
algorithm for \grkrri.

\subsubsection{Termination}

Obviously, the deletion of constraints in $S$ makes a new proof of termination
necessary, since the proof of Lemma~\ref{lem:optimised-alg-terminates} relied
on this fact. Please note, that the Lemma~\ref{lem:path-is-poly-long}  still
holds for \grkrri.

\begin{lemma}[Termination]
\label{lem:grkrri-alg-terminates}
Any sequence of rule applications starting from a \CS $S = \{ x_0 \models
\phi \}$ of the \grkrri algorithm is finite.
\end{lemma}

\begin{proof}
  The sequence of rule applications induces a sequence of trees. As before,
  the depth and out-degree of this tree is bounded in $|\phi|$ by
  Lemma~\ref{lem:path-is-poly-long}. For each variable $x$, $\Lab(x)$ is a
  subset of the finite set $\sub(\phi)$. Each application of a rule either
  \begin{itemize}
  \item adds a constraint of the form $x \models \psi$ and hence adds an element
    to $\mathcal{L}(x)$, or
  \item adds fresh variables to $S$ and hence adds additional nodes to the
    tree $G(S)$, or
  \item adds a constraint to a node $y$ and deletes all subtrees rooted at
    successors of $y$.
  \end{itemize}
  
  Assume that algorithm does not terminate. Due to the mentioned facts this
  can only be because of an infinite number of deletions of subtrees.  Each
  node can of course only be deleted once, but the successors of a single node
  may be deleted several times. The root of the completion tree cannot be
  deleted because it has no predecessor.  Hence there are nodes which are
  never deleted.  Choose one of these nodes $y$ with maximum distance from the
  root, i.e., which has a maximum number of ancestors in $\prec_S$.
  Suppose that $y$'s successors are deleted only finitely many times. This can
  not be the case because, after the last deletion of $y$'s successors, the
  ``new'' successors were never deleted and thus $y$ would not have maximum
  distance from the root. Hence $y$ triggers the deletion of its successors
  infinitely many times. However, the $\rulechoose$-rule is the only rule
  that leads to a deletion, and it simultaneously leads to an increase of
  $\Lab(y)$, namely by the missing concept which caused the deletion of $y$'s
  successors.  This implies the existence of an infinitely increasing chain of
  subsets of $\sub(\phi)$, which is clearly impossible. \qed
\end{proof}

\subsubsection{Soundness and Completeness}

\begin{lemma}[Soundness]
  \label{lem:grkrri-soundness}
  Let $\phi$ be a \grkrri-formula in NNF. If the completion rules can be
  applied to $\{ x_0 \models \phi \}$ such that they yield a complete and
  clash-free c.s., then $\phi \in \satgrkrri$.
\end{lemma}

\begin{proof}
  Let $S$ be a complete and clash-free \CS obtained by a sequence of rule
  applications from $\{ x_0 \models \phi \}$. We show that the canonical
  structure $\M_S$ is indeed a model of $\phi$, where the canonical structure for
  \grkrri is defined as in Definition~\ref{def:canonical-structure}. Please
  note, that we need the condition ``$Rxy \in S$ iff $R^{-1}yx \in S$'' to
  make sure that all information from the \CS is reflected in the canonical
  structure.
  
  By induction over the norm of formulae $\| \psi \|$ as defined in the proof
  of Lemma~\ref{lem:soundness-optimised-algorithm}, we show that, for a
  complete and clash-free \CS $S$, $x \models \psi \in S$ implies $\M_S,x
  \models \psi$. The only interesting cases are when $\psi$ starts with a
  modal operator.
  \begin{itemize}
  \item $x \models \modgeq \omega n \psi \in S$ implies $\omega^S(x,\psi) \geq
    n$ because $S$ is complete. Hence, there are $n$ distinct variables $y_1,
    \dots, y_n$ with $y_i \models \psi \in S$ and $Rxy_i \in S$ for each $1 \leq
    i \leq n$ and $R \in \omega$. By induction, we have $\M_S, y_i \models
    \psi$ and $(x,y_i) \in \omega^{\M_S}$ and hence $\M_S,x \models \modgeq
    \omega n \psi$.
  \item $x \models \modleq \omega n \psi \in S$ implies, for any $R\in \omega$
    and any $y$ with $Rxy \in S$, $y \models \psi \in S$ or $y \models \nneg
    \psi \in S$. For any predecessor of $x$, this is guaranteed by the
    $\rulechoose$-rule, for any successor of $x$ by the $\rulegeq$-rule which
    is suspended until no non-generating rule rules can applied to $x$ or any
    predecessor of $x$ together with the reset-restart mechanism that is
    triggered by constraints ``moving upwards'' from a variable to its
    predecessor.
  
    We show that $\sharp \omega^{\M_S}(x,\psi) \leqslant \sharp
    \omega^S(x,\psi)$: assume $\sharp \omega^{\M_S}(x,\psi) > \sharp
    \omega^S(x,\psi)$.  This implies the existence of some $y$ with $(x,y) \in
    R^{\M_S}$ for each $R \in \omega$ and $\M_S, y \models \psi$ but $y \models
    \psi \not\in S$.  This implies $y \models \nneg \psi \in S$, which, by
    induction yields $\M_S, y \models \nneg \psi$ in contradiction to $\M_S,y
    \models \psi$. 
  \end{itemize}
  
  Since constraints for the initial variable $x_0$ are never deleted from $S$,
  we have that $ x_0 \models \phi \in S$ and hence $\M_S,x_0 \models \phi$ and 
  $\phi \in \satgrkrri$.
  \qed
\end{proof}

The following lemma combines the local and global completeness proof for the
\grkrri-algorithm

\begin{lemma}[Completeness]
  \label{lem:grkrri-completeness}
  If $\phi \in \satgrkrri$ in NNF, then there is a sequence of the
  \grkrri-rule starting with $S = \{ x_0 \models \phi \}$ that results in a
  complete and clash-free c.s.
\end{lemma}

\begin{proof}
  Let $\M$ be a model for $\psi$ and $\overline{\mathcal{R}}_\phi$ the set of
  relations that occur in $\phi$ together with their inverse. We use $\M$ to
  guide the application of the non-deterministic completion rules by
  incremently defining a function $\alpha$ mapping variables from the \CS to
  elements of $W^\M$. The function $\alpha$ will always satisfy the following
  conditions:
  \[
  \left. 
    \begin{array}{@{}l@{\;}l@{}}
      1. & \text{if } x \models \psi \in S  \text{ then } \M,\alpha(x) \models \psi\\
      2. & \text{if } Rxy \in S \text{ then } \{ R \mid Rxy \in S \} = \{ R \mid
      (\alpha(x),\alpha(y)) \in R^\M \} \cap \overline{\mathcal{R}}_\phi\\ 
      3. & \text{if } y,z \text{ are distinct variables such that } \{R_1 xy,
      R_2 xz \} \subseteq S, \text{ then } \alpha(y) \neq \alpha(z)
    \end{array}
  \right \} \; (*)
  \]

  \noindent \textsc{Claim:} Whenever $(*)$ holds for a \CS $S$ and a function
  $\alpha$ and a rule is applicable to $S$ then it can be applied in a way that
  maintains $(*)$.

  \begin{itemize}
  \item The $\ruleand$-rule: if $x \models \psi_1 \wedge \psi_2 \in S$, then
    $\M,\alpha(x)\models (\psi_1 \wedge \psi_2)$.  This implies $\M,\alpha(x)
    \models \psi_i$ for $i=1,2$, and hence the rule can be applied without
    violating $(*)$.
  \item The $\ruleor$-rule: if $x \models \psi_1 \vee \psi_2 \in S$, then
    $\M,\alpha(x) \models (\psi_1 \vee \psi_2)$. This implies $\M,\alpha(x)
    \models \psi_1$ or $\M,\alpha(x) \models \psi_2$. Hence the $\ruleor$-rule
    can add a constraint $x \models \chi$ with $\chi \in \{ \psi_1, \psi_2 \}$
    such that $(*)$ still holds.
  \item The $\rulechoose$-rule: obviously, either $\M,\alpha(y) \models \psi$ or
    $\M,\alpha(y) \models \nneg \psi$ for any variable $y$ in $S$.  Hence, the
    rule can always be applied in a way that maintains $(*)$.  Deletion of
    nodes does not violate $(*)$.
  \item The $\rulegeq$-rule: if $x \models \modgeq \omega n \phi' \in S$, then
    $\M, \alpha(x) \models \modgeq  \omega n \phi'$. This implies $\sharp
    \omega^\M(\alpha(x),\phi') \geqslant n$.  We claim that there is an element $t
    \in W^\M$ such that
    \[
    \left. 
      \begin{array}{l}
        (\alpha(x),t) \in R^\M \text{ for each } R \in \omega, \text{ and } \M, t \models
        \psi, \text{ and }\\  
        t \not \in \{ \alpha(y) \mid Rxy \in S \}
      \end{array}
      \quad 
    \right \} (**)
    \]
    We will come back to this claim later. Let $\psi_1, \dots, \psi_k$ be an
    enumeration of the set $\{\psi \mid x \models \langle \sigma
    \rangle_{\bowtie m} \in S \}$  The $\rulegeq$-rule can add the constraints
    \begin{align*}
      S' & =  \{ y \models \psi_i \mid \M,t \models
      \psi_i \} \cup \{ y \models \nneg \psi_i \mid \M,t \not\models \psi_i \}\\
      S'' & = \{ Rxy \mid R \in \overline{\mathcal{R}}_\phi, (\alpha(x),t) \in
      R^\M \} \cup \{ Ryx \mid R \in \overline{\mathcal{R}}_\phi, (t,\alpha(x)) \in
      R^\M \}
    \end{align*}
    as well as $\{ y \models \phi' \}$ to $S$. If we set $\alpha' := \alpha[y \mapsto
    t]$, then the obtained \CS together with $\alpha'$ satisfies $(*)$.
  
    Why does there exists an element $t$ that satisfies $(**)$? Let $s \in
    W^\M$ be an arbitrary element with $(\alpha(x),s) \in \omega^\M$ and $\M,s
    \models \psi$ that appears as an image of an arbitrary element $y$ with
    $Rxy \in S$ for some $R \in \overline{\mathcal{R}}_\phi$.  Condition 2 of
    $(*)$ implies that $Rxy \in S$ for any $R \in \omega$ and also $y \models
    \psi \in S$ must hold as follows:
  
    Assume $y \models \psi \not \in S$.  This implies $y \models \nneg \psi
    \in S$: either $y \prec_S x$, then in order for the $\rulegeq$-rule to be
    applicable, no non-generating rules and especially the $\rulechoose$-rule
    is not applicable to $x$ and its ancestor, which implies $\{ y \models
    \psi, y \models \nneg \psi \} \cap S \neq \emptyset$. If not $y \prec_S x$
    then $y$ must have been generated by an application of the
    $\rulegeq$-rule to $x$. In order for this rule to be applicable no
    non-generating rule may have been applicable to $x$ or any of its
    ancestors. This implies that at the time of the generation of $y$ already
    $x \models \modgeq \omega n \psi \in S$ held and hence the $\rulegeq$-rule
    ensures $\{y \models \psi,y \models \nneg \psi\} \cap S \neq \emptyset$.
    
    In any case $y \models \nneg \psi \in S$ holds and together with Condition
    1 of $(*)$ this implies $\M,s \not \models \psi$ which contradicts $\M,s
    \models  \psi$.

    Together this implies that, whenever an element $s$ with $(\alpha(x),s) \in
    \omega^\M$ and $\M,s \models \psi$ is assigned to a variable $y$ with $Rxy 
    \in S$, then it must be assigned to a variable that contributes to $\sharp 
    \omega^S(x,\psi)$. Since the $\rulegeq$-rule is applicable there are less
    than $n$ such variables and hence there must be an unassigned element
    $t$ as required by $(**)$.
  \end{itemize}
  
  This concludes the proof of the claim. The claim yields the lemma as
  follows: obviously, $(*)$ holds for the initial \CS $\{ x_0 \models \phi
  \}$, if we set $\alpha(x_0):= s_0$ for an element $s_0$ with $\M, s_0
  \models \phi$ (such an element must exist because $\M$ is a model for
  $\phi$). The claim implies that, whenever a rule is applicable, then it can
  be applied in a manner that maintains $(*)$.
  Lemma~\ref{lem:grkrri-alg-terminates} yields that each sequence of rule
  applications must terminate, and also each \CS  for which $(*)$ holds is
  necessarily clash-free. It cannot contain a clash of the form $\{x \models
  p, x \models \neg p \} \subseteq S$ because this would imply $\M,\alpha(x)
  \models p$ and $\M, \alpha(x) \not \models p$.  It can neither contain a
  clash of the form $x \models \modleq \omega n \psi \in S$ and $\sharp
  \omega^S(x,\psi) > n$ because $\alpha$ is an injective function on $\{ y
  \mid Rxy \in S \}$ and preserves all relations in
  $\overline{\mathcal{R}}_\phi$. Hence $\sharp \omega^S(x,\psi) > n$ implies
  $\sharp \omega^\M(x,\psi) > n$, which cannot be the case since $\M,
  \alpha(x) \models \modleq \omega n \psi$.  \qed
\end{proof}

As a corollary of Lemma~\ref{lem:grkrri-alg-terminates},
\ref{lem:grkrri-soundness}, and \ref{lem:grkrri-completeness} we get:

\begin{corollary}
  The \grkrri-algorithm is a non-deterministic decision procedure for
  \satgrkrri.
\end{corollary}

\subsection{Complexity of the Algorithm}

As for the optimised algorithm for \grkr, we have to show that the
\grkrri-algorithm can be implemented in a way that consumes only polynomial
space. This is done similarly to the \grkr-case, but we have to deal with two
additional problems: we have to find a way to implement the ``reset-restart''
caused by the $\rulechoose$-rule, and we have to store the values of the
relevant counters $\omega^S(x,\psi)$. It is impossible to store the values for
each possible intersection of relations $\omega$ because the are exponentially
many of these. Fortunately, storing only the values for those $\omega$ which
actually appear in $\phi$ is sufficient.

\begin{lemma}
  \label{lem:grkrri-algo-in-pspace}
  The \grkrri-algorithm can be implemented in \pspace.
\end{lemma}

\begin{proof}
  Consider the algorithm in Figure~\ref{fig:grkrri-decision-proc}, where
  $\Omega_\phi$ denotes all intersections of relations that occur in
  $\phi$. As the algorithm for \grkr, it re-uses the space used to check for
  the existence of a complete and clash-free ``subtree'' for each successor
  $y$ of a variable $x$. Counter variables are used to keep track of the
  values $\sharp \omega^S(x,\psi)$ for all relevant $\omega$ and $\psi$. This
  can be done in polynomial space. Resetting a node and restarting the
  generation of its successors is achieved by resetting all successor
  counters. Please note, how the predecessor of a node is taken into account
  when initialising the counter variables.
  
  Since the length of paths in a \CS is polynomial bounded in $|\phi|$ and
  all necessary book-keeping information can be stored in polynomial space,
  this proves the lemma. \qed
\end{proof}

\begin{figure}
  \begin{center}
    \leavevmode
    \begin{tabbing}
      \small
      \hspace{2em}\=\hspace{2em}\=\hspace{2em}\= \kill
      $\grkrri-\textsc{SAT}(\phi) := \texttt{sat}(x_0,\{x_0 \models \phi \})$\\
      $\texttt{sat}(x,S)$: \\
      \> allocate counters  $\sharp \omega^S(x,\psi)$ for all $\omega \in
      \Omega_\phi$ and $\psi \in \sub(\phi)$. \\
      \hspace{5pt} \textsf{restart:}\\
      \> \texttt{for each} counter $\sharp \omega^S(x,\psi)$:\\
      \>\> \texttt{if} $x$ has a predecessor  $y \prec_S x$ \texttt{and}
      $\omega \subseteq \{ R \mid Rxy \in S \}$ \texttt{and} $y \models \psi
      \in S$\\
      \>\>\> \texttt{then} $\sharp \omega^S(x,\psi) := 1$ \texttt{else} $\sharp \omega^S(x,\psi):=0$\\
      \> \texttt{while} (the $\ruleand$- or the $\ruleor$-rule can be
      applied at $x$) \texttt{and}  (S is clash-free) \texttt{do}\\
      \>\> apply the $\ruleand$- or the $\ruleor$-rule to $S$.\\
      \> \texttt{od}\\
      \> \texttt{if} $S$ contains a clash \texttt{then}
      \texttt{return} ``not satisfiable''.\\
      \> \texttt{if} the $\rulechoose$-rule is applicable to the constraint
      $x \models \langle \omega \rangle_{\bowtie n} \psi \in S$\\
      \>\> \texttt{then return} ``restart with $\psi$''\\
      \> \texttt{while} (the $\rulegeq$-rule applies to a constraint $x \models 
      \modgeq \omega n \phi' \in S$) \texttt{do} \\
      \>\> $S_{\text{new}} := \{ y \models \phi' \} \cup S' \cup S''$\\
      \>\> \texttt{where}\\
      \>\>\> $y$ is a fresh variable\\
      \>\>\> $\{ \psi_1, \dots, \psi_k \} = \{  \psi \mid x \models \langle \sigma
      \rangle_{\bowtie m} \psi \in S \}$\\
      \>\>\> $S' = \{y \models \chi_1, \dots, y \models \chi_k \}$, and\\
      \>\>\> $\chi_i$ is chosen non-deterministically from $\{ \psi_i, \nneg
      \psi_i \}$ \\
      \>\>\> $S'' = \{ R_1xy, R_1^{-1}yx, \dots, R_lxy,R^{-1}_lyx \}$\\
      \>\>\> $\{R_1, \dots, R_l \}$ is chosen non-deterministically with
      $\omega \subseteq \{R_1, \dots, R_l \} \subseteq \overline{\mathcal{R}}_\phi$\\
      \>\> \texttt{for each} $\psi$ with $y \models \psi \in S'$
      \texttt{and} $\sigma \in \Omega_\phi$ with $\sigma \subseteq \{ R \mid Rxy
      \in S'' \}$ \texttt{do}\\
      \>\>\> increment $\sharp \sigma^S(x,\psi)$\\
      \>\> \texttt{if} $x \models \modleq \sigma m \psi \in S$ and $\sharp
      \sigma^S(x,\psi) > m$ \\
      \>\>\> \texttt{then return} ``not satisfiable''.\\
      \>\> $result := \texttt{sat}(y,S \cup S_{\text{new}})$\\
      \>\> \texttt{if} $result =$ ``not satisfiable'' \texttt{then return}
      ``not satisfiable''\\
      \>\> \texttt{if} $result =$ ``restart with $\psi$'' \texttt{then} \\
      \>\>\> $S := S \cup \{ x \models \chi \}$\\
      \>\>\> \texttt{where} $\chi$ is chose non-deterministically from $\{\psi,
      \nneg \psi \}$ \\
      \>\>\> \texttt{goto} \textsf{restart}\\
      \> \texttt{od} \\
      \> remove the counters for $x$ from memory.\\
      \> \texttt{return} ``satisfiable''
    \end{tabbing}
    \caption{A  non-deterministic \pspace  decision procedure for \satgrkrri.}
    \label{fig:grkrri-decision-proc}
  \end{center}
\end{figure}

Obviously, \satgrkrri{} is \pspace-hard, hence
Lemma~\ref{lem:grkrri-algo-in-pspace} and Savitch's
Theorem~\cite{JCSS::Savitch1970} yield:

\begin{theorem}
  Satisfiability for \grkrri is \pspace-complete if the numbers in the input
  are represented using binary coding.
\end{theorem}

As a simple corollary, we get the solution of an open problem
in~\cite{IC::DoniniLNN1997}:

\begin{corollary}
  Satisfiability for $\mathcal{ALCNR}$ is \pspace-complete if the numbers in
  the input are represented using binary coding.
\end{corollary}

\begin{proof}
  The DL $\mathcal{ALCNR}$ is a syntactic restriction of the DL
  $\mathcal{ALCQIR}$, which, in turn, is a syntactical variant of
  \grkrri. Hence, the \grkrri-algorithm can immediately be applied to
  $\mathcal{ALCNR}$-concepts. \qed
\end{proof}



\section{Conclusion}

We have shown that by employing a space efficient tableaux algorithm
satisfiability of the logic \grkr can be decided in \pspace, which is an optimal result
with respect to worst-case complexity. Moreover, we have extended the
technique to the logic \grkrri, which extends \grkr both by inverse relations
and intersection of relations. This logic is a notational variant of the
Description Logic $\ALCQIR$, for which the complexity of concept
satisfiability has also been open. This settles the complexity of the DL
$\mathcal{ALCNR}$ for which the upper complexity bound with binary coding had
also been an open problem~\cite{IC::DoniniLNN1997}.  While the algorithms
presented in this work certainly are only optimal from the viewpoint of
worst-case complexity, they are relatively simple and will serve as the
starting-point for a number of optimisations leading to more practical
implementations. They also serve as tools to establish the upper complexity
bound of the problems and thus shows that tableaux based reasoning for \grkr
and \grkrri can be done with optimum worst-case complexity.  This establishes
a kind of ``theoretical benchmark'' that all algorithmic approaches can be
measured against.

\subsubsection*{Acknowledgments.}

I would like to thank Franz Baader, Ulrike Sattler, and an anonymous referee for
valuable comments and suggestions. Part of this work was supported by the DFG, 
Project No. GR 1324/3-1.



\begin{thebibliography}{DLNN97}

\footnotesize

\setlength{\parskip}{0pt}

\bibitem[AvBN98]{AndrekaVanBenthemNemeti:JPL:98}
H.~Andr\'eka, J.~van~Benthem, and I.~N\'emeti
\newblock Modal languages and bounded fragments of predicate logic.
\newblock {\em Journal of Philosophical Logic}, 27(3):217--274, 1998.

\bibitem[BBH96]{Baader96a}
F.~Baader, M.~Buchheit, and B.~Hollunder.
\newblock Cardinality restrictions on concepts.
\newblock {\em Artificial Intelligence}, 88(1--2):195--213, 1996.

\bibitem[CLN94]{CLN94}
D. Calvanese, M. Lenzerini, and D. Nardi.
\newblock A Unified Framework for Class Based Representation  Formalisms.
\newblock {\em Proc. of {KR}-94}, 1994.

\bibitem[dHR95]{JLOGC::HoekR1995}
W. Van der Hoek and M. De Rijke.
\newblock Counting objects.
\newblock {\em Journal of Logic and Computation}, 5(3):325--345, June 1995.

\bibitem[DLNN97]{IC::DoniniLNN1997}
F.~M. Donini, M. Lenzerini, D. Nardi, and W. Nutt.
\newblock The complexity of concept languages.
\newblock {\em Information and Computation}, 134(1):1--58, 10~April 1997.

\bibitem[Fin72]{Fine72}
K.~Fine.
\newblock In so many possible worlds.
\newblock {\em Notre Dame Journal of Formal Logic}, 13:516--520, 1972.

\bibitem[GS96]{Giunchiglia96a}
F.~Giunchiglia and R.~Sebastiani.
\newblock Building decision procedures for modal logics from propositional
  decision procedures---the case study of modal {K}.
\newblock {\em Proc. of {CADE}-13}, LNCS 1104. Springer, 1996.

\bibitem[HB91]{HollunderBaader-KR-91}
B.~Hollunder and F.~Baader.
\newblock Qualifying number restrictions in concept languages.
\newblock In {\em Proc. of {KR}-91}, pages
  335--346, Boston (USA), 1991.

\bibitem[HM92]{HalpernMoses92}
J.~Y. Halpern and Y.~Moses.
\newblock A guide to completeness and complexity for model logics of knowledge
  and belief.
\newblock {\em Artificial Intelligence}, 54(3):319--379, April 1992.

\bibitem[HS97]{Hustadt97b}
U.~Hustadt and R.~A. Schmidt.
\newblock On evaluating decision procedures for modal logic.
\newblock In {\em Proc. of IJCAI-97)}, volume~1, pages 202--207, 1997.

\bibitem[HST99]{HoSatTo-LPAR-99}
I.~Horrocks, U.~Sattler, and S.~Tobies.
\newblock Practical Reasoning for Expressive Description Logics.
\newblock In H.~Ganzinger and A.~Voronkov, editors, {\em Proceedings of the
  6th International Conference on Logic for Programming and Automated
  Reasoning {(LPAR'99)}}
\newblock LNAI number 1705, Springer-Verlag.

\bibitem[Lad77]{SICOMP::Ladner1977}
R.~E. Ladner.
\newblock The computational complexity of provability in systems of modal
  propositional logic.
\newblock {\em SIAM Journal on Computing}, 6(3):467--480, September 1977.

\bibitem[OS97]{JLOGC::OhlbachS1997}
H.~J. Ohlbach and R.~A. Schmidt.
\newblock Functional translation and second-order frame properties of modal
  logics.
\newblock {\em Journal of Logic and Computation}, 7(5):581--603, October 1997.

\bibitem[OSH96]{OhlbachSchmidtHustadt96}
H.~J. Ohlbach, R.~A. Schmidt, and U.~Hustadt.
\newblock Translating graded modalities into predicate logic.
\newblock In H.~Wansing, editor, {\em Proof Theory of Modal Logic}, volume~2 of
  {\em Applied Logic Series}, pages 253--291. Kluwer, 1996.

\bibitem[Sav70]{JCSS::Savitch1970}
W.~J. Savitch.
\newblock Relationships between nondeterministic and deterministic tape
  complexities.
\newblock {\em Journal of Computer and System Sciences}, 4(2):177--192, April
  1970.

\bibitem[Sch91]{Schild91a}
K.~Schild.
\newblock A correspondence theory for terminological logics: Preliminary
  report.
\newblock In {\em Proc. of IJCAI-91}, pages 466--471, 1991.

\bibitem[Sch97]{Schmidt97a}
R.~A.~Schmidt.
\newblock Resolution is a decision procedure for many propositional modal
  logics: Extended abstract.
\newblock In M.~Kracht, M.~de~Rijke, H.~Wansing, and M.~Zakharyaschev, editors,
  {\em Advances in Modal Logic '96}. CLSI Publications, 1997.

\bibitem[SSS91]{Schmidt-Schauss91}
M.~Schmidt-Schau{\ss} and G.~Smolka.
\newblock Attributive concept descriptions with complements.
\newblock {\em Artificial Intelligence}, 48:1--26, 1991.


\bibitem[Tob99]{Tobies-CADE-99}
S.~Tobies.
\newblock A {PSpace} algorithm for graded modal logic.
\newblock In H.~Ganzinger, editor, {\em Automated Deduction -- CADE-16, 16th
  International Conference on Automated Deduction}, LNAI 1632, pages 52--66,
  Trento, Italy, July~7--10, 1999. Springer-Verlag.

\end{thebibliography}
\end{document}